\documentclass[nofootinbib,aps]{revtex4}

\usepackage{ulem}

\usepackage{amstext,amsmath,amssymb,bbm}
\usepackage[dvips]{graphicx,epsfig}
\usepackage{latexsym}

\usepackage{color}

\usepackage{psfrag}
\newcommand{\Ref}[1]{(\ref{#1})}

\newcommand{\N}{\mathbb{N}}

\def\be{\begin{equation}}
\def\ee{\end{equation}}
\def\bes{\begin{eqnarray}}
\def\ees{\end{eqnarray}}

\def\arr{\rightarrow}

\def\6{\langle}
\def\9{\rangle}

\def\f{\frac}

\def\wt{\widetilde}

\def \vphi{\varphi}
\newcommand{\SU}{\mathrm{SU}}

\def\hh{{\cal H}}
\def\co{{\cal O}}

\def\dj{{}^j d}
\def\hgf{{}_2F_1}

\def\pp{\partial}

\def\6{\langle}
\def\9{\rangle}

\def\half{\mbox{$1\over2$}}
\def\1{{{\mathbbm 1}}}

\def\pp{\partial}
\def\rr{{\cal R}}

\begin{document}
\title{Bulk Entropy in Loop Quantum Gravity}
\author{{\bf Etera R. Livine}\footnote{etera.livine@ens-lyon.fr}}
\affiliation{Laboratoire de Physique ENS Lyon, CNRS UMR 5672, 46 All\'ee d'Italie, 69364 Lyon Cedex 07, France}
\author{{\bf Daniel R. Terno}\footnote{dterno@physics.mq.edu.au} }
\affiliation{Centre for Quantum Computer Technology,
Department of Physics, Macquarie University, Sydney NSW 2109, Australia}

\begin{abstract}
\begin{center} {\small ABSTRACT} \end{center}

In the framework of loop quantum gravity (LQG), having quantum black
holes in mind, we generalize the previous boundary state counting
(gr-qc/0508085) to a full bulk state counting. After a suitable
gauge fixing we are able to compute the bulk entropy of a bounded
region (the ``black hole") with fixed boundary. This allows us to
study the relationship between the entropy and the boundary area in
details and we identify the holographic regime of LQG where the
leading order of the entropy scales with the area. We show that in
this regime we can fine tune the factor between entropy and area
without changing the Immirzi parameter.

\end{abstract}
\maketitle


%
%
%
%

\section{Introduction}

To understand the deep structure of  Loop Quantum Gravity (LQG)
\cite{lqg}, in particular the holographic principle and quantum black
holes, it is necessary to analyze in detail the entropy counting.
Most of the work on black hole entropy in LQG  focuses on boundary
state counting, especially in the isolated horizon framework
\cite{isoh}. In the present work we propose to extend these
considerations to the bulk entropy.

We use the framework outlined in \cite{ourbh}. We consider a given spin network state for the 3d
geometry of the (canonical) hypersurface and focus on an arbitrary bounded region. The previous
work \cite{ourbh} analyzed the boundary entropy of such a region assuming a totally mixed state,
i.e having no knowledge about its interior. We naturally recovered the area-entropy law. In the
present work, we extend this calculation to the bulk entropy. More precisely, we retain the
information about the graph (underlying the quantum geometry state) inside the considered region
and, under given boundary conditions, count all the distinct spin network states on that graph.
Then the entropy -- defined as the logarithm of the number of states-- depends on the topology of
the graph through its number of loops.

In the trivial topology case with no loop, we recover the previous result obtained by counting
boundary states \cite{ourbh}. As soon as the topology becomes non-trivial the entropy diverges.
Nevertheless, we identify a symmetry responsible for this divergence and  get a finite entropy
after suitable gauge fixing. The entropy increases with the number of loops. For a complicated
enough graph topology we find that the entropy can grow arbitrarily large compared to the boundary
area. However, we also identify a regime for which the entropy still scales as the area. We call it
the ``holographic regime". There it turns out to be possible to arbitrarily adjust the
proportionality factor between the entropy and the area (without changing the Immirzi parameter).
If quantum gravity is to be a holographic theory, then the LQG dynamics should be such that
projecting on physical states selects this regime.

Finally, the gauge fixing that we use seems to be related to a gauge fixing of the Hamiltonian
constraint, but more investigations are required to understand that relationship.

\section{Graph Topology and State Counting}

Let us start with an arbitrary spin network state on the canonical
hypersurface $\Sigma$.  Consider a connected bounded region ${\cal
R}$ of its graph\footnote{It is usually assumed that the graph is
locally finite, i.e that the valency of each vertex is finite.} that
includes a finite set of vertices and the edges that connect them.
The boundary $\pp\rr$ is the set of edges which have only one end
vertex laying in $\rr$. We can picture $\rr$ as a 3-ball and
$\pp\rr$ as its boundary 2-sphere punctured by the boundary edges.

\begin{figure}[htbp]
\begin{center}
\includegraphics[width=5cm]{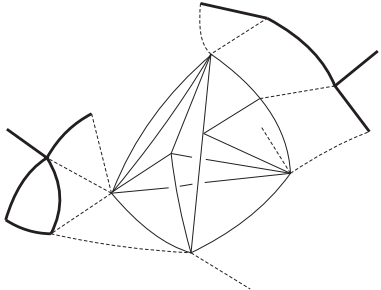}
\end{center}
\caption{\label{figbig}
\small{
The internal edges   are shown as regular lines, the boundary edges
are dashed, and the selected exterior edges are rendered in bald.
}}
\end{figure}

Let us call $\Gamma$ the graph inside $\rr$ and assume that the
boundary $\pp\rr$ is made of $n$ edges. The spin network state
carries spin labels ($\SU(2)$ representations) attached to each edge
of the graph. In particular, the boundary data consists of the spin
labels $j_1,..j_n$ of the $n$ boundary edges puncturing $\pp\rr$. We
assume these  labels fixed once and  for all as defining the
boundary. We further take them all equal to the lowest spin
$j_1=..=j_n=1/2$. As it was shown in
\cite{ourbh}, this hypothesis simplifies the calculations and it is
straightforward to generalize the calculations to any generic
configuration. For such a choice of spins the number of boundary
edges $n$ is necessarily even, and we  therefore assume in the
following that the boundary is made of $2n$ edges.

In  \cite{ourbh} we motivated defining boundary states as
intertwiners between the $2n$ boundary representations --- or
equivalently as singlet states in the tensor product of the $2n$
spin $j_1,..j_{2n}$. Counting the dimension $N_\pp$ of the
intertwiner space\footnotemark~gives  the (boundary) entropy $S_\pp$
in term of the binomial coefficient $C_{2n}^n$:
\be
S_\pp\,\equiv\, \log N_\pp \,=\, \log \f{1}{n+1}{2n\choose n}
\underset{n\arr\infty}{\sim} 2n\log2 -\f32\log n +\dots
\ee
Assuming that every $j=1/2$ puncture contributes a microscopic area
$a_{\f12}l_P^2$ in Planck units, we find that the entropy satisfies
the usual  proportionality law at the leading order, and has a
$-3/2$ logarithmic correction:
\be
S_\pp \,\sim\, \f{A}{l_P^2}\,\f{\log 2}{a_{\f12}}\,-\f32\log A+\dots ,
\ee
where the total boundary area is $A=2na_{\f12}l_P^2$. In the
standard LQG framework, the microscopic area is given in terms of
the Immirzi parameter $a_{\f12}=\gamma\sqrt{3}/2$. Then the
semi-classical relation $S\sim A/4l_P^2$  can be used to fix the
value of the Immirzi ambiguity $\gamma$. The isolated horizon
framework presents a different though similar calculation for the
entropy. It leads to a different ratio $S/A$ and thus a different
value of $\gamma$, but also to a different value of the logarithmic
correction - usually $\f{1}2$ instead of the present $\f{3}2$.

\footnotetext{
In \cite{ourbh}, we assumed the black hole state defined by the totally mixed state $\rho\propto\1$
on the intertwiner space. This is naturally the state seen by the external observer who does not
have any information on the internal black hole state. It is also a static state, which does not
evolve under unitary evolution, and thus corresponds to the physical set-up of a black hole. Then
the entropy of this state is of course the logarithm of the intertwiner space dimension.}

This boundary entropy calculation can be interpreted as computing the number of spin network states
assuming that the graph $\Gamma$ inside the region $\rr$ is reduced to a single vertex. Assuming
that the outside observer  have no access to any information about the graph inside $\rr$, we have
indeed coarse-grained this graph to a single point: this can be dubbed the ``black point" model.

\medskip

In the present work we investigate the effects of keeping a
non-trivial graph $\Gamma$ on the entropy calculation. As a
spin-network wave functional is a function of the (gravitational)
holonomies, it is possible to have an impression that degrees of
freedom are attached to the edges of $\Gamma$. However, due to the
gauge invariance, degrees of freedom are truly carried by the loops
of the graph.

Consider $\Gamma$ with $V$ vertices, $E$ internal edges and $E_\pp=2n$ external legs. Then the
number of independent loops is $L=E-V+1$. When   $\Gamma$ is a tree, $L=0$,  the entropy  is
exactly the same as for the trivial graph with a single vertex. As soon as the graph possesses a
non-trivial topology\footnote{Let us point out that the graph topology does not a priori have any
relation to the topology of the spatial hypersurface. There are two points of view about the
manifold topology. It is either assumed right at the start and  we consider embedded graphs.
Otherwise we consider abstract graphs, containing only combinatorial and algebraic data, and the
hypersurface topology is an emerging semi-classical notion. We favor the latter point of view.},
$L\ge 1$, we  obtain an infinite number of states and thus of degrees of freedom. Indeed, loops of
a spin network can carry arbitrary spins $j\in\N/2$ independently of each other. To overcome this
obstacle, we note that this divergence is associated to a further gauge invariance. This gauge
invariance is generated by the action of the holonomy operators on the loops of the spin network
state. To count physical degrees of freedom and to get a meaningful finite entropy, we gauge fix
this action
\cite{petal}. We use the simplest gauge fixing and fix the spin
associated to each loop of $\Gamma$. It actually requires fixing the
spin of only one link of each loop. Having done this, we will see in
the following section that we indeed obtain a finite dimensional
Hilbert space. The resulting entropy does not depend of the choice
of the spins used in the gauge fixing and only depends on the
size/area of the boundary $E_\pp=2n$ and on the complexity of the
graph $\Gamma$ described by the number of loops $L$.

This gauge fixing of the holonomy operators on the graph loops can
be related to the gauge fixing of the Hamiltonian constraint. In the
context of the topological BF theory, this would actually exactly
coincide with the action of the Hamiltonian constraint\footnote{The
action of the holomony operators in the BF theory generates the
translational symmetry on the B field. It needs to be dealt with in
order to get the physical degrees of freedom of the theory}. In the
case of LQG, it is more subtle  and we should investigate further
the validity and physical interpretation of our gauge fixing. While
we postpone this for future work, we insist that in our framework
this gauge fixing is natural. First, we gauge fix the holonomies on
loops inside the considered region $\rr$. This a priori does not
affect the dynamics of the exterior of $\rr$. On the other hand, if
we did not gauge fix them we would be over-counting the number of
states (seen by an external observer). What would be non-trivial is
to gauge fix the action of holonomies on loops crossing the
boundary/horizon $\pp\rr$. Second, we do check that the number of
degrees of freedom does not depend on the gauge fixing and that the
action of the holonomies creates isomorphic copies of the same
``physical" Hilbert space. In short, we are disregarding the
internal excitations that do not couple to the exterior of the
region and that would produce an overcounting in the entropy
ascribed to $\rr$ by an external observer.

In the following sections we study in detail the one-loop case and
then show that the calculations can be straightforwardly generalized
to an arbitrary graph topology. Finally we discuss the different
regimes of entropy corresponding to the different scaling of the
number of loops $L$ with the size of the boundary $n$.

\section{The one-loop case}

Let us start by reviewing the black-point case, $L=0$. The Hilbert
space of spin networks of a tree is isomorphic to the space of
intertwiners between the boundary representations
\cite{ourbh}. Its dimension is given by:
\be
\dim \hh_0 \,=\,
\int_{\SU(2)} dg\, \chi_{\f12}(g)^{2n}\,=\,
\f{1}{n+1}{2n\choose n},
\ee
where $dg$ is the normalized Haar measure on the $\SU(2)$ Lie group
and $\chi_{1/2}(g)$ is the character in the fundamental spin-$1/2$
representation. This formula can be interpreted in terms of a random
walk with a mirror in origin. The binomial coefficients $C_{2n}^n$
give the  number of returns to the origin for the usual random walk,
while the mirror introduces the factor $1/(n+1)$.

\begin{figure}[t]
\psfrag{j}{$j$}
\psfrag{k}{$k$}
\psfrag{s}{$\f12$}
\begin{center}
\includegraphics[width=7cm]{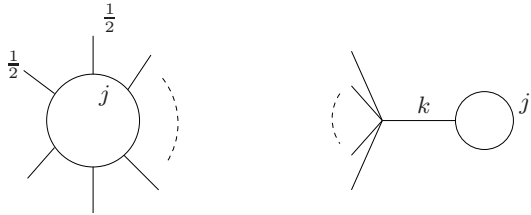}
\end{center}
\caption{\label{oneloop}
\small{
One-loop spin networks with the $2n$ boundary links: the internal loop carries the spin $j$.}}
\end{figure}

Consider now the one-loop diagram on the left of Fig.~\ref{oneloop}.
It is a loop to which $2n$ legs are attached, so all the vertices
are 3-valent\footnote{An arbitrary one-loop graph is equivalent to
this form due to the $\SU(2)$ gauge invariance at every vertex (see
e.g.
\cite{petal}.}. Pick an arbitrary link $e_0$ and fix the spin which it carries, $j_{e_0}=j$.
This obviously fixes the action of the holonomy operator on the
loop: acting with a holonomy carrying a spin $J$ on this state would
change the spin-$j$ representation into the tensor product
representation $j\otimes J$.

Assume that $j\ge n/2$. Moving from the initial link $e_0$ to the next link $e_1$, we see that
$e_1$ can carry either the spin $j-1/2$ or $j+1/2$ since the external leg between $e_0$ and $e_1$
injects a spin $1/2$ into the diagram. Going on one finally arrives back to the initial link $e_0$
and the initial representation $j$. Therefore, the number of possible spin labeling
$j,j\pm\f12,\dots,j\pm\f12$ is exactly the number of returns to the origin of a random walk after
$2n$ iterations:
\be
\dim \hh_1={2n\choose n}.
\ee
We get a finite result which is different from the tree case. Moreover, it does not depend on the
chosen spin $j$. This Hilbert space leads to the following asymptotic behavior of the entropy:
\be
S_1\underset{n\arr\infty}{\sim} 2n\log 2-\f12\log n+\dots
\ee
It does not affect the leading order proportional to the area but only the logarithmic correction.

There is nevertheless a subtlety: the assumption that the spin $j$ is large enough compared to the
boundary size $n$. Indeed if $j$ is smaller than $n/2$, there exists the possibility that the
random walk along the loop will ascribe a spin 0 to a link. In such a case, we can not move down to
a $-1/2$ spin but can only go back up to $+1/2$. Then we get a smaller Hilbert space. We discard
this case because we do not consider it as a true one-loop graph anymore. Indeed, if a link is
labeled with a spin $0$, it is just as if that link did not exist since the corresponding spin
network wave function would not depend on the holonomy on that link. Then if we remove that link
from the graph, we end up with a tree again. Therefore, we interpret this situation when $j\le n/2$
as describing a  superposition of a 0-loop and 1-loop graphs. However, we can still compute exactly
the dimension of the Hilbert space. For this purpose, it is more convenient to use the another
one-loop diagram that is shown on the right of Fig.~\ref{oneloop}.

We now look at the one-loop diagram with two vertices: all the
boundary links merge into a single link to which the loop is then
attached. Assume that the loop still carries a fixed spin $j$ and we
call $k$ the spin carried by the intermediate link. On one side, we
have an intertwiner between the $2n$ spin $1/2$ and the spin $k$. As
long as $k\le n$, the dimension of this intertwiner space is
\cite{ourbh}:
\be
d^{(n)}_k\,=\,{2n\choose n+k}-{2n\choose n+k+1}=\frac{2k+1}{n+k+1}{2n\choose n+k}.
\ee
On the other side, we have a unique 3-valent intertwiner between two representations $j$ and the
same representation $k$. The corresponding intertwiner space is of dimension one as long as $k\le
2j$. The total number of spin network states amounts to summing over all possible $k$'s from $0$ to
the maximal allowed spin $k_{max}\equiv \min(n,2j)$:
\be
N_1^{(j)}\,=\,
\sum_{k=0}^{k_{max}}d^{(n)}_k
\,=\,
{2n\choose n}-{2n\choose n+ k_{max}+1}.
\ee
When $j$ is larger than $n/2$, we recover the previous result with $N_1^{(j)}=N_1={2n\choose n}$.
At the other end of the spectrum, when $j$ vanishes, we recover the 0-loop case with
$N_1^{(j)}=N_0={2n\choose n}-{2n\choose n+1}$. Finally, we see that the values of the spin $j$
carried by the loop between $0$ and $n/2$ interpolate between the tree case and the 1-loop case.
This allows us to interpret this intermediate situation as a superposition of the 0-loop and 1-loop
cases, and moreover to identify the true 1-loop case as the $j\ge n/2$ regime when the entropy does
not depend anymore on the specific value of $j$.

To conclude this section, we showed that considering a one-loop graph over a tree increases the
entropy. After carefully gauge fixing, we obtain a finite entropy which differs from the 0-loop
case only by the logarithm correction, $-1/2$ instead of $-3/2$.

In the following sections, we generalize this analysis to an arbitrary number of loops. We will
show that if the number of loops is fixed it will only affect the logarithm correction while if we
allow the number of loops to scale with $n$ it may well also affect the leading order.


\section{The two-loop case}

\begin{figure}[htbp]
\psfrag{ja}{$j_a$}
\psfrag{jb}{$j_b$}
\psfrag{k}{$k$}
\psfrag{s}{$\f12$}
\begin{center}
\includegraphics[width=7cm]{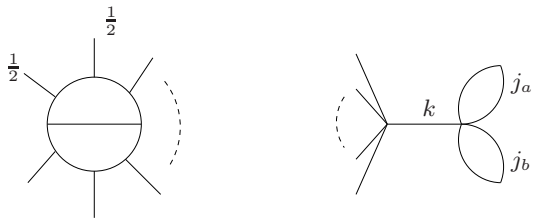}
\end{center}
\caption{\label{LoopToPetal}
\small{
Two-loop graphs with boundary links}}
\end{figure}

  We now move to the entropy counting in the two-loop case. We work
with the graph shown on fig.\ref{LoopToPetal} and we fix the spins
along the two loops to the same value $j$ for the sake of
simplicity. We further assume as previously that $j\ge n/2$ so that
we consider a pure two-loop graph and not a superposition with a
0-loop or 1-loop configuration. The dimension of the Hilbert space
of spin networks on $\Gamma$ is given by the following integral:
$$
N_2\,=\,
\int dg\,\chi_{\f12}(g)^{2n}\,\chi_j(g)^4.
$$

For computational purpose, it is convenient to express as a sum over
the intermediate spin label $k$. It is straightforward to obtain
\be
N_2\,=\,\sum_{k=0}^n d_k^{(n)}\,\left[(2j+1)(2k+1)-\f32k(k+1)\right].
\ee
The two terms can be computed exactly\footnotemark:
\be
\sum_{k=0}^n (2k+1)d_k^{(n)}=2^{2n},\qquad
\sum_{k=0}^n k(k+1)d_k^{(n)}=n{2n \choose n} \sim \sqrt{n}2^{2n}.
\ee

\footnotetext{A useful identity for the degeneracy coefficients is
$$
\sum_{k=0}^n d^{(n)}_k \,\sin (2k+1)\theta
\,=\,
2^{2n}\,\sin\theta\cos^{2n}\theta,
$$
which is derived by computing the trace of $\SU(2)$ group elements in the representation
$(\f12)^\otimes{2n}$. Differentiating this equation, we obtain the following formula for the
polynomial averages over the $d_k^{(n)}$ distribution, for $l\in\N$:
$$
\sum_{k=0}^n (2k+1)^{2l+1}d_k^{(n)}
\,=\, (-1)^l2^{2n}\partial_\theta[\sin\theta\cos^{2n}\theta]_{\theta=0}.
$$}

As we see we have a residual $j$-dependent term, which actually diverges as $j$ grows to infinity.
Neglecting\footnote{The second term in $k(k+1)$ is indeed be neglected if $j$ goes to infinity
faster than $\sqrt{n}$. This condition is automatically satisfied since we assume that $j\ge n/2$.}
the extra-term $k(k+1)$, the $j$-dependence can be factorized. Therefore, we interpret this as a
symmetry that we haven't gauge fixed yet. Indeed we have gauge fixed the action of the holonomy
operator along the loops $a$ and $b$ by requiring $j_a=j_b=j$, but the action of the holonomy along
the loop $a\cup b$ is not yet fixed. This leaves a freedom in the intertwiner which leads to that
$(2j+1)$ factor, resulting in a over-counting in the Hilbert space dimension.

\begin{figure}[htbp]
\psfrag{j}{$j$}
\psfrag{J}{$J$}
\psfrag{k}{$k$}
\begin{center}
\includegraphics[width=34mm]{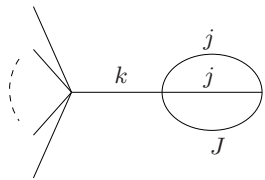}
\end{center}
\caption{\label{twoloops}
\small{Fully gauge-fixed two-loop spin network: we fix the spins carried by the three edges on the loops.}}
\end{figure}

This can easily be seen considering the two-loop diagram as shown in Fig.~\ref{twoloops}. $J$ can
still vary from 0 to $(2j+1)$. This is the freedom that requires gauge-fixing. It is possible to
compute the intertwiner dimension for different values of $J$. It is possible to compute the
intertwiner dimension for different values of $J$. For instance, we get:
$$
N_2^{(J=j)}=\sum_{k=0}^n (2k+1)d_k^{(n)},
\qquad N_2^{(J=2j)}=\sum_{k=0}^n (k+1) d_k^{(n)}.
$$
Both these examples, have the same behavior up a factor with a leading order given by $\sum_k k
d_k^{(n)}$. They give the same entropy in the asymptotic limit $n\arr+\infty$ (same leading order
and logarithmic correction):
\be
S_2\sim \log 2^{2n} \sim 2n \log 2 +\dots
\ee
with a vanishing logarithmic correction (they are of course  sublog corrections). We see that
adding one loop from the one-loop case corresponds to implementing a $+\f12$ factor in the
logarithmic correction without affecting the leading order proportional to the area. We generalize
this statement to arbitrary number of loops in the following section.

\section{Gauge Fixing and Generic Bulk Entropy}

\subsection{Informal Arguments}

For a generic graph with $L$ loops, following the one-loop and
two-loop cases, it is reasonable to expect the gauge-fixed dimension
at leading order to scale as
\be
N_L\sim \sum_{k=0}^n k^{L-1}d_k^{(n)}.
\label{guess}
\ee
Assuming that $L$ is fixed, the asymptotics is in the large $n$ limit only changes the logarithmic
correction of the entropy:
\be
S_L\,\equiv\log N_L\,\sim 2n\log 2 +\left(\f L 2 -1\right)\log n +\dots
\label{asymptguess}
\ee
Using the same techniques as in \cite{ourbh}, it is straightforward to approximate the sum over $k$
by an integral and then compute the asymptotics of $N_L$ by a saddle point approximation. Then only
the logarithmic correction changes by a factor $+\f12$ with each loop because the distribution
$d_k^{(n)}$ looks like a Gaussian peaked at $k=0$ with a width scaling as $\sqrt{n}$.

More precisely, denoting $x\equiv k/n \in [0,1]$, we can approximate the dimensions $d_k^{(n)}$ for
large $n$ using the Stirling formula:
\be
d_k^{(n)}\,\sim\, \f2{\sqrt\pi}\,
\f{2^{2n}}{\sqrt n}\f{x}{(1+x)\sqrt{1-x^2}}\,e^{-n\vphi(x)},
\ee
with the exponent $\vphi(x)$ given by:
\be
\vphi(x)\,=\, (1+x)\log(1+x)+(1-x)\log(1-x).
\ee
Therefore, the full Hilbert space dimension can be approximated by an integral:
$$
N_L\,\sim\,\f2{\sqrt\pi}\,2^{2n}n^{L-\f12}\int_0^1 dx\, \f{x^L}{(1+x)\sqrt{1-x^2}}\,e^{-n\vphi(x)}.
$$
The exponent $\vphi(x)$ is always positive for $x\in[0,1]$. It has a unique fixed point at $x=0$.
For large $n$, we can use the Gaussian approximation, $\vphi(x)=x^2+{\cal O}(x^3)$. Then we need to
evaluate $\int_0^\infty dx\, x^L\exp(-nx^2)$, which scales as $n^{-(L+1)/2}$. Finally, this leads
to the  asymptotics for the entropy given above in \Ref{asymptguess}.

So far we have kept the number of loops $L$ fixed as the boundary area $n$ was taken to infinity.
This number does not affect the leading order of the entropy, but only changes the pre-factor of
the logarithmic correction. However, we see that if we allow the graph complexity to grow with the
boundary size, we are able to change the leading order behavior of the entropy and  change the
proportionality factor between the entropy $S$ and the area $n$.

Below we make this argument more precise and compute the dimension and entropy exactly for a
specific choice of gauge-fixing. We find a missing factor $1/(L-1)!$ in the guessed dimension
\Ref{guess}. This changes the asymptotical behavior when $L$ is allowed to be arbitrarily large.
Then we keep the asymptotics described above for fixed $L$ while allowing the loop number $L$ to
scale as $n$ changes the leading order factor\footnotemark~between the $S_L$ and $n$.

\footnotetext{
Assuming that the number of loops goes as $L=\alpha n$ with a fixed ratio $\alpha>0$, it is
straightforward to extract the asymptotics of the entropy:
$$
S_\alpha\,\equiv\,\log \left[\f{1}{(L-1)!}\sum_{k=0}^n k^{L-1}d_k^{(n)}\right]
\,\sim\,\lambda n-\f12\log n+\dots,
$$
where the precise value of $\lambda>0$ depends on $\alpha$ but is generically different from $2\log
2$. The key to this calculation is that the weight $k^{L-1}$ now contributes to the exponent which
get modified to $\wt{\vphi}(x)=\vphi(x)-\alpha\log x$. The fixed point $x_0$ of $\wt{\vphi}$ is not
$x=0$ anymore. In particular $\wt{\vphi}(x_0)\ne 0$ and this leads to a term proportional to $n$ in
the Gaussian approximation.}

\subsection{The Entropy Formula}

\begin{figure}[b]
\psfrag{j}{$j$}
\psfrag{j2}{$2j$}
\psfrag{j4}{$4j$}
\psfrag{jL}{$2^{L-1}j$}
\psfrag{k}{$k$}
\begin{center}
\includegraphics[width=40mm]{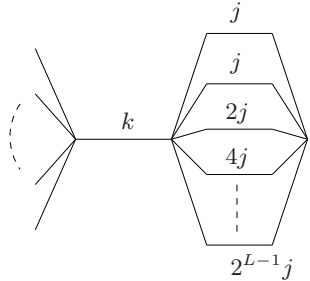}
\end{center}
\caption{\label{manyloops}
\small{Gauge-fixed spin networks with $L$ loops: the $L+1$ edges of $\Gamma$ carry the labels $j,j,2j,\ldots,2^{L-1}j$.}
}
\end{figure}

Let us now consider the $L$-loop graph with three vertices $A,B,C$ as shown on Fig~\ref{manyloops}.
The $2n$ boundary links combine into intermediate spin $k$ at the vertex $A$. On the other side,
the $L$ loops combine at the vertex $B$ to that same intermediate link. Finally the intertwiner at
the $C$ node closes the graph and describes how the $L$ loops are coupled to each other.

Our ``gauge fixing", or more precisely, the choice of a sector of the spin network Hilbert space,
is achieved by  fixing the spin labels on the $L+1$ edges defining the $L$ loops. We label them
sequentially by representations $j,j,2j,4j,8j,..,2^{L-1}j$. As long as $j$ is assumed larger than
$n/2$ as before, this leads to an entropy which does not depend on the gauge-fixing representation
$j$ but only on the number of loops $L$ and the number of boundary edges $n$.

We need to underline that we are not only fixing the action of the
holonomy operator along the $L$ loops. Indeed, our specific choice
of representation labels is not only convenient for computational
purposes but also it fully fixes the unique intertwiner at the node
$C$. This amounts to fixing $L-2$ representation labels within the
node $C$. This means that we are truly fixing $(L+1)+(L-2)=2L-1$
quantum numbers, and not only $L$ as we first expected. We are in
fact counting the number of possible intertwiners at the vertices
$A$ and $B$. At a na\"\i ve level, the entropy that we compute
counts only the different ways that the internal loops couple to the
external links while we disregard the number of ways that the loops
couple to each other by fixing the intertwiner at the internal node
$C$. At a mathematical level, if one fixes less representation
labels than we do (less than $2L-1$), one gets an entropy which
depends on $j$ and diverges as $j$ grows large. This is a symptom of
insufficient gauge-fixing. Nevertheless, there remains the open
issue of understanding the physical interpretation and relevance of
such a gauge-fixing. In particular, what is the precise symmetry
that we are gauge-fixing? We leave this question for future
investigation. However, even without such an interpretation in term
of symmetry and gauge-fixing, we still have identified sectors of
the spin network Hilbert space with different asymptotic behaviors
of the entropy of the considered region which only depend on the
number of loops of the graph (supporting the spin network states)
inside that region.

To compute the entropy, we need to find the number of intertwiners
at the vertices $A$ and $B$. At the vertex $A$ we have the same
degeneracy $d_k^{(n)}$ as previously. As for the vertex $B$, having
$L$ loops leads to the following degeneracies for the spin-$k$
subspaces as long as $k\leq 2^{L-1}j$ for $L\ge 1$:
\be
\dj^{(L)}_k={k+L-1\choose k}.
\ee
This formula is straightforward to prove  by induction using the identity $\sum_{k=0}^K
{k+L-1\choose k}= {K+L\choose K}$. As a result, the dimension of the spin network space inside ${\cal R}$ is
\be
N_L=\sum_{k=0}^n d^{(n)}_k \dj^{(L)}_k,\label{ndef}
\ee
where we have assumed $n\le 2j$. For $L\geq 2$ this can be slightly simplified\footnotemark:
\be
N_L\,=\,\sum_{k=0}^n d^{(n)}_k {k+L-1\choose k}
\,=\,\sum_{k=0}^n {2n\choose n+k}{k+L-2\choose k},\label{nicesum}
\ee
\footnotetext{This sum has an explicit closed form in terms of hypergeometric
functions,
\be
N_L\,
=\frac{(2n)!}{(n!)^2}\,\hgf(-n, L-1, 1+n;-1).\label{hyperbullshit}
\ee}
For a small number of loops\footnotemark, we easily get the exact
results for $N_L$, using properties of the binomial coefficients
\cite{concrete}. For $L=1$, we recover the previous result
$N_1=\sum_k d^{(n)}_k ={2n\choose n}$, with an asymptotic expression for the entropy $S_1\sim
2n\log2 -\f12
\log n$. For $L=2$ one obtains
\be
N_2=\sum_{k=0}^n{2n\choose n+k}=2^{2n-1}+\frac{1}{2}{2n\choose n},
\ee
with the entropy $S_2$ having no logarithmic correction for large $n$ (of course, it still contains
sub-logarithmic corrections).
For $L=3$, we can also compute:
\be
N_3=\sum_{k=0}^n{2n\choose n+k}(k+1)=\frac{n+1}{2}{2n\choose n}+2^{2n-1},
\ee
with a logarithmic correction being $+\f12$.

\footnotetext{To reach higher values of $L$, the number of states $N_L$ can be calculated by induction with the help
of the following recurrence relation:
$$
\sum_{k=0}^n k^m {2n\choose n+k}=\sum_k k^{m-2}[n^2-(n^2-k^2)]{2n\choose n+k}=
n^2 \sum_k k^{m-2}{2n\choose n+k}-2n(2n-1)\sum_k k^{m-2}{2n-2\choose n-1+k}.
$$}


\subsection{The Entropy Asymptotics}

In the analysis of  the entropy asymptotics, we distinguish between three cases. First, if the
number of loops $L$ is held fixed (or more generally $L$ stays negligible compared to $n$), we
recover the results given above: the leading order $2n\log 2$ does not change, while the
non-trivial topology of the graph affects the logarithmic correction which becomes $(L/2-1)\log n$.

In the second case  $L$ is scaling as $n$: the entropy depends on the limit of the ratio
$\alpha\equiv L/n$ as $n$ goes to infinity. In the third case  $L$ grows very large compared to
$n$: if the ratio $L/n$ is unbounded, the leading order of the entropy drastically changes and will
not scale  proportionally to $n$ anymore.

The asymptotic expression for all the above cases can be derived from a saddle point approximation
of the sum $N_L=\sum_k{2n \choose n+k}{k+L-2 \choose k}$ as shown in details in appendix. It also
allows to obtain the sub-leading terms with an excellent precision.

When $L\ll n$ we get a compact expression,
\be
S_L=\log N_L, \qquad
N_L\sim\frac{2^{2n-L-3}n^{L/2-1}}{\Gamma(L/2)},\label{lsmall}
\ee
where $\Gamma(z)=(z-1)!$. This still gives a good estimate of the sub-leading terms: for example,
when $L=7$ and $n=10000$, the difference between the exact and asymptotic values is $\Delta
S=0.354794$, while $S\approx 1.39\times10^4$.

\begin{figure}[t]
\begin{center}
\includegraphics[width=70mm,angle=-90]{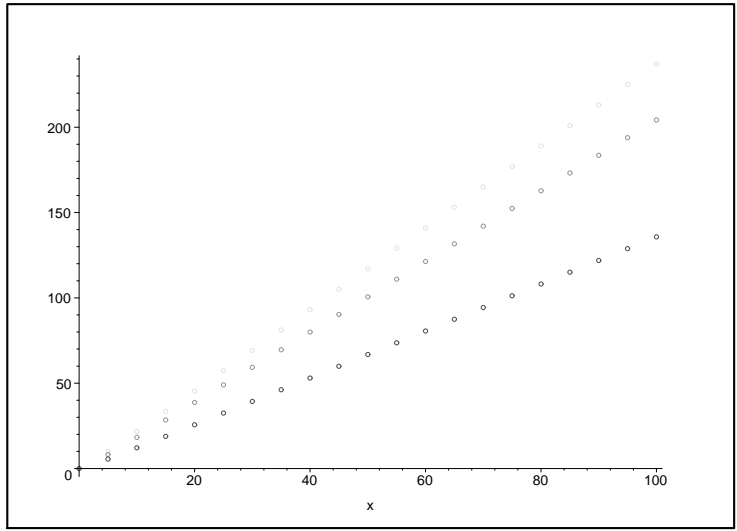}
\end{center}
\caption{\label{l0l1l2}
\small{Plot of the entropy $S_L(n)$ for $L=0$, $L=n$ and $L=2n$
for the number of punctures $n$ running from 0 to 100. The leading behavior is linear in $n$ but
the slope depends on the scaling $L/n$. The numerics give $\lambda\sim1.386$ for the boundary
entropy $L=0$, while we get $\lambda\sim2.074$ for $L=n$ and $\lambda\sim 2.400$ for $L=2n$. These
numerical values fit the analytical result (up to two decimals) derived in appendix.} }
\end{figure}

When $L=\alpha n$ with a fixed ratio $\alpha$, we can still apply the same saddle point technique.
Using Stirling formula, we approximate the dimension $N_L$ by the integral expression,
\be
N_L\,\sim\,
\f{2^{2n}n^{L-\f32}}{\sqrt{\pi}(L-2)!e^{L-2}}\,
\int_0^1 dx\,\f{1}{\sqrt{x}\sqrt{1-x^2}(x+\alpha)^{\f32}}\,e^{-n\wt{\vphi}(x)},
\ee
with the new $\alpha$-dependent exponent,
\be
\wt{\vphi}(x)\,=\, \vphi(x)+x\log x -(x+\alpha)\log(x+\alpha).
\ee
As shown in details in appendix, the Gaussian approximation controls the asymptotic behavior of the
entropy:
\be
S_{L=\alpha n}\,\sim\,
\lambda n -\f12\log n +\dots
\ee
The important new result is that the leading order factor $\lambda\sim S/n$ depends on $\alpha$ and
is not fixed  to the standard factor $2\log 2$ anymore. We show on Fig.~\ref{l0l1l2} a plot with
Maple numerics for the entropy $S_L(n)$ for $L=0$ (as a reference) and $\alpha=1,2$. While the
leading behavior is always linear, the slop $\lambda\,\equiv\,S/n$ clearly depends on the value of
the ratio $\alpha=L/n$.

The final case is when the complexity of the graph $L$ grows much
faster than the boundary size $n$. In this case, the entropy grows
faster than linearly compared to the boundary area:
\be
S_{L\gg n}\,\sim\, n(\log L-\log n) +n -\f12\log n+\dots
\ee

\section{The Regimes of Bulk Entropy}

After a suitable gauge fixing, we have managed to compute the finite bulk entropy counting the
number of spin network states supported by a fixed graph $\Gamma$ inside the considered bounded
region ${\cal R}$ with fixed boundary conditions. The entropy $S$ only depends on the number of
links $n$ puncturing the boundary $\pp{\cal R}$ and the number of loops $L$ of $\Gamma$ which
quantifies the complexity of the graph. Computing the asymptotics of $S_L$, we distinguish three
regimes.
\begin{itemize}

\item {\bf The logarithmic regime~:}\\
The number of loops $L$ is held fixed while $n$ is take to infinity. The leading order of $S_L$ is
linear in $n$ and is exactly the same as for the boundary entropy \cite{ourbh}. The complexity of
the graph only affects the factor in front of the logarithmic correction $\log n$, which increases
with $L$. Therefore, if a (quantum) black hole is belongs to this regime, fixing the smallest spin
$j=\f12$ fixes the Immirzi parameter.

\item {\bf The holographic regime~:}\\
The number of loops $L$ scales proportionally to the horizon area $n$. The entropy still grows
linearly in $n$ in the leading order but the proportionality factor changes and depends on the
ratio $\alpha=L/n$. This is a sector of the spin network space where the entropy still scales with
the boundary area and not faster. If the full LQG theory is to be (strongly) holographic, then the
dynamics should restrict the physical states to this regime: the graph complexity can not grow
faster than the boundary size. On the other hand, even if the LQG theory is not restricted to this
regime, a quantum black hole state should lay in this regime if we want to respect the
semi-classical area-entropy law. Then since the factor $S/n$ depends on the ratio $\alpha$, we can
adjust this new parameter $\alpha$ so to get $S\sim A/4$ without changing neither the Immirzi
parameter nor the area quantum $a_{\f12}$. In this scenario, the black hole entropy calculation
does not fix the Immirzi parameter. From another perspective, we can say that the factor $\alpha$
renormalizes the Immirzi parameter $\gamma$. Finally we need the dynamics to select the physical
sector of the spin network space and provide us with the ``right" value of $\alpha$. Of course,
this scenario only holds if we use the bulk entropy and do not restrict ourselves to work with the
boundary entropy.

\item {\bf The non-linear regime~:}\\
If the number of loops grows much faster than $n$, i.e $L/n\arr\infty$, then the entropy is free to
grow non-linearly with $n$. Since the leading order is  $n\log L$ , we can, e. g., get an entropy
scaling with the ``volume" $n^{3/2}$ for an exponential growth of the type $L\sim 2^{\sqrt{n}}$. Of
course, one should keep in mind that the volume of ${\cal R}$ actually depends on the bulk state
and does not always scale as $n^{3/2}$. It could well be larger (or smaller) if the space is
tightly curved.

\end{itemize}


\section*{Conclusions}

We explored the notion of ``bulk entropy" in the framework of loop
quantum gravity, in order to generalize the calculations of boundary
entropy of \cite{ourbh}. Our aim was to compute the number of (spin
network) states describing the quantum geometry of a bounded region
of space, living on a fixed graph and with fixed boundary
conditions. This bulk entropy is obviously related to the complexity
of the graph and was found to be related to the number of loops $L$
of that graph. The straightforward calculation then gives at first
an infinite entropy. Nevertheless, after fixing a certain number of
labels of the spin network states, we were able to extract
meaningful finite results. This lead to the identification of
different sectors of the spin network Hilbert space depending on $L$
with different asymptotical behavior of the entropy.

We can a priori have an entropy growing as large as we want if we
choose a number of loops $L$ large enough. Nevertheless, two sectors
are particularly interesting. If the graph complexity is fixed while
the boundary area grows, then the bulk entropy $S$ has the same
leading order than the boundary entropy but the logarithmic
correction to the entropy changes and increases with $L$. On the
other hand, if $L$ grows linearly with the boundary area $A$, then
the bulk entropy also grows linearly with the area but the ratio
$S/A$ depends explicitly on the ratio $L/A$. We call this sector the
``holographic regime".

If we want black holes to satisfy the area-entropy law, the quantum
black hole state must necessarily be in the holographic regime.
However, this regime is generic enough to allow for a fine-tuning of
the ratio $S/A$ to $1/4$ without changing the Immirzi parameter (i.e
the value of the minimal quantum of area). Indeed, we can always
change the area-complexity ratio $L/A$ to adjust the value of $S/A$.
Our point of view is that it is the LQG dynamic that is supposed to
select the physical regime and value of $L/A$.

There are two important issues to address in this  scenario. First, should the considered entropy
for the space region be its boundary entropy or its bulk entropy? We see a priori no reason why the
degrees of freedom inside the region would not couple to the space outside. Then the bulk degrees
of freedom would affect the black hole evaporation. However, the case of a black hole might
(should?) be drastically different from a generic region of space. This is a question that needs to
be addressed by the LQG dynamics. The second issue is directly related to our approach. Our entropy
calculation relies on a partial fixing of the spin network labels. We discussed that it could be
interpreted as a gauge fixing of the Hamiltonian constraint, but this needs to be worked out in
details. Thus, the limits of validity of the present work depend on understanding the legitimacy of
this gauge fixing. However, even if it does not turn out to be interpreted as the gauge fixing of a
certain symmetry, we have nevertheless identified the different sectors of the spin network
kinematical space with different bulk entropy behavior, but we still need to provide these
different sectors with a proper physical interpretation.

\appendix

\section{Computing the Asymptotics of the Entropy}
The asymptotic expressions for $n\rightarrow\infty$ and the different regimes of $L=L(n)$ are based
on the use of Stirling formula, replacement of the sum by the (Euler-MacLaurent) integral and the
saddle point (Gaussian) approximation. It is convenient for the analysis to introduce a shifted
loop number $l=L-2$ and consider the ratio $\alpha\equiv\,l/n$.

The first factor in the sum of Eq.~(\ref{nicesum}) takes the form
\be
{2n \choose n+k}\sim\frac{2^{2n}}{\sqrt{\pi n}}f(k/n)e^{-n\vphi(k/n)},
\ee
where
\be
f(x)=\frac{1}{\sqrt{1-x^2}}, \qquad \vphi(x)= (1+x)\log(1+x)+(1-x)\log(1-x).
\ee
The second factor becomes
\be
{k+l\choose k}\sim\frac{1}{e^l
l!}\sqrt{\frac{(k+l)}{k}}\frac{(k+l)^{l+k}}{k^k}=\frac{n^l}{e^l l!}g(k/n,l/n)e^{-n\psi(k/n,l/n)},
\ee
where
\be
g(x,\alpha)=\sqrt{\frac{(x+\alpha)}{x}}, \qquad
\psi(x,\alpha)=x\log x -(x+\alpha)\log(x+\alpha).
\ee
The sum of Eq.~(\ref{nicesum}) is replaced by the integral
\be
N_l\sim \frac{2^{2n}}{\sqrt{\pi n}}\frac{n^{l+1}}{e^l l!}
\int_0^1 dx\, f(x)g(x,\alpha)e^{-n\wt{\vphi}(x,\alpha)},
\ee
with $\wt{\vphi}=\vphi(x)+\psi(x,\alpha)$. The saddle point approximation requires the knowledge of
the  derivatives of the exponent:
\be
\pp_x\wt{\vphi}=\log\f{x(1+x)}{(x+\alpha)(1-x)},\qquad
\pp_x^2\wt{\vphi}=\frac{\alpha+2\alpha x+(2-\alpha)x^2}{(x+\alpha)(x-x^3)}.
\ee
The unique fixed point, $\pp\wt{\vphi}(x_0)=0$, in the interval $[0,1]$ is
\be
x_0\,\equiv\,\f{1}{4}(\sqrt{\alpha^2+8\alpha}-\alpha),
\ee
and it is easy to check that $\pp^2\wt{\vphi}$ is always positive on $[0,1]$.

In particular, for $0<\alpha<\infty$ the minimum of $\wt{\vphi}$ is inside the interval and we use
the Gaussian approximation
$$
\int_0^1 dx\,y(x)\,e^{-n\wt{\vphi}(x)}
\,\sim\,
y(x_0)\sqrt{\f{2\pi}{n\pp^2\wt{\vphi}(x_0)}}\,e^{-n\wt{\vphi}(x_0)},
$$
that gives the following asymptotics for the entropy:
\be
S(l,n)=\log N_l, \qquad
N_l\sim\frac{2^{2n+1/2}n^l}{e^ll!}f(x_0)g(x_0,\alpha)\frac{e^{-n\wt{\vphi}(x_0,\alpha)}}{\sqrt{\pp^2\wt{\vphi}(x_0,\alpha)}}.
\label{gauss}
\ee

Hence the leading order term of the entropy
\be
S(l,n)\sim\lambda n-\half\log n,
\ee
is given in terms of $\alpha$:
\bes
\lambda&\equiv&
2\log 2 -\wt{\vphi}(x_0)-\alpha\log\alpha, \\
&=& 5\log 2 -\alpha\log 4\alpha
+\alpha\log[3\alpha+\sqrt{\alpha(\alpha+8)}]-\log[8-\alpha(4+\alpha-\sqrt{\alpha(\alpha+8)})],
\nonumber
\ees
while the subleading terms have a more cumbersome appearance.

As a matter of fact, this approximation is in an excellent agreement with the exact results: e. g.,
$S(l=10,n=10000)\approx 1.39\times10^4$  and the error is $\Delta S\approx-0.0077$, increasing only
to $\Delta S(l=10,n=40000)\approx-0.0080$. At the other extreme, $S(l=1.5\times 10^5,n=1000)\approx
6022.7$ with $\Delta S\approx-8.3\times 10^{-5}$.


When $\alpha$ runs from 0 to $\infty$, the fixed point $x_0$ varies from 0 to 1. In the special
case $\alpha=1$ when $l=n$,
$$
x_0=\f12,\quad \wt{\vphi}(x_0)=-\log 2,\quad\pp^2\wt{\vphi}(x_0)=4,
$$
and we get simple asymptotics:
\be
S(l=n)\,\sim\, 3n\log 2 -\f12\log n-\f12\log\pi.
\ee
In this case, we can actually give a faster proof of the asymptotics:
$$
N_{l=n}\,=\,
\sum_{k=0}^n{2n \choose n+k}{n+k \choose k}
\,=\,
\sum_k \f{(2n)!}{(n!)^2}{n \choose k}
\,=\,
\f{(2n)!}{(n!)^2}2^n.
$$
In the limit regime  $l\ll n$, we have
$$
x_0\underset{\alpha\arr0}{\sim} \sqrt{\f\alpha 2}\,,\quad
\wt{\vphi}(x_0)\sim \half \alpha(\log 2-1-\log \alpha),\quad
\pp^2\wt{\vphi}(x_0)\sim 4-\sqrt{2 \alpha}+5\alpha/2,
$$
which improves the estimate of Eq.~(\ref{lsmall}) to
\bes
S(l\ll n)& = & 2n\log2 + \frac{l}{2}\log n - \frac{l}{2} \log l + (1 - \log2) \frac{l}{2}
 - \frac{1}{2}\log l
-\frac{1}{2}\log(4\pi)
\\  &  & - \frac{1}{12 l} + \frac{1}{\sqrt{2}}l\sqrt{\frac{l}{n}}+\frac{5}{4\sqrt{2}}\sqrt{\frac{l}{n}}
 - \frac{l^2}{4n} - \frac{9}{32}\frac{l}{n}+\co\left(\frac{1}{n^2}\right). \nonumber
\ees
Finally, in the  case that $l\gg n$, we have
$$
x_0\underset{\alpha\arr\infty}{\sim} 1-\f2\alpha\,,\quad
\wt{\vphi}(x_0)\sim -(\alpha+1)\log \alpha+(2\log 2-1)-\frac{5}{\alpha},\quad
\pp^2\wt{\vphi}(x_0)\sim \frac{\alpha}{2}+\frac{7}{2}-\frac{1}{2\alpha},
$$
and Eq.~(\ref{gauss}) yields a simple expression for the entropy,
\be
S(l\gg n)\sim n\log l-(n+\half)\log n+n-\half\log2\pi+(6n-1)/12l+\co(n^2/l^2).
\ee


\end{document}